\documentclass{elsart}
\usepackage{epsfig}

\usepackage{amssymb}

\begin{document}

\begin{frontmatter}

% Title, authors and addresses

% use the thanksref command within \title, \author or \address for footnotes;
% use the corauthref command within \author for corresponding author footnotes;
% use the ead command for the email address,
% and the form \ead[url] for the home page:
% \title{Title\thanksref{label1}}
% \thanks[label1]{}
% \author{Name\corauthref{cor1}\thanksref{label2}}
% \ead{email address}
% \ead[url]{home page}
% \thanks[label2]{}
% \corauth[cor1]{}
% \address{Address\thanksref{label3}}
% \thanks[label3]{}

 \title{A~Coupled~Cavity~Micro~Fluidic~Dye~Ring~Laser}
\author{M. Gersborg-Hansen, S. Balslev, N.~A. Mortensen, and A. Kristensen}
 \address{MIC -- Department of Micro and Nanotechnology,\\
Technical University of Denmark,\\ Building 345 East, \O rsteds
Plads, DK-2800 Kongens Lyngby, Denmark}

\begin{abstract}
We present a laterally emitting, coupled cavity micro fluidic dye
ring laser, suitable for integration into lab-on-a-chip micro
systems. The micro-fluidic laser has been successfully designed,
fabricated, characterized and modelled. The resonator is formed by
a micro-fluidic channel bounded by two isosceles triangle mirrors.
 The micro-fluidic laser structure is defined using photo
lithography in 10~$\rm{\mu m}$ thick SU-8 polymer on a glass
substrate. The micro fluidic channel is sealed by a glass lid,
using PMMA adhesive bonding. The laser is characterized using the
laser dye Rhodamine 6G dissolved in ethanol or ethylene glycol as
the active gain medium, which is pumped through the micro-fluidic
channel and laser resonator. The dye laser is optically pumped
normal to the chip plane at 532 nm by a pulsed, frequency doubled
Nd:YAG laser and lasing is observed with a threshold pump pulse
energy flux of around 55 ${\rm \mu J/mm^2}$. The lasing is
multi-mode, and the laser has switchable output coupling into an
integrated polymer planar waveguide. Tuning of the lasing
wavelength is feasible by changing the dye/solvent properties.
\end{abstract}

\end{frontmatter}

\section{Introduction}

Integrable lab-on-a-chip light sources are essential for on-chip
spectral analysis of chemical samples \cite{verpoorte2003}. For
these applications dye lasers are of particular interest due to
the possibility of tuning the wavelength in the visible range.
Micro-fluidic dye lasers have recently been demonstrated by glass
\cite{cheng2004} and polymer \cite{helbo2003a} micro-fabrication
and tunability has also been reported \cite{bilenberg2003a}.
Polymer-based laterally emitting single-mode dye-lasers were
reported recently \cite{balslev2004b}. The latter implementation
is advantageous due to the direct integrability with lab-on-a-chip
systems without additional hybridization steps
\cite{balslev2004a}.

In this paper we present a new type of polymer-based laterally
emitting micro-fluidic dye laser utilizing coupled cavities.
Lasing is achieved with a new cavity design (see
Fig.~\ref{fig:sample}) relying on total internal reflection at the
interface between the polymer (SU-8 with refractive index
$n_1=1.59$) and the surrounding air at an incidence angle of
$45^\circ$. Bleaching of the dye is avoided by a regenerating flow
through the micro-fluidic channel. In this way, the dye may also
be dynamically changed, enabling real-time tunability of the
laser. Furthermore, the output coupling is switchable.

In the following sections we describe the resonator structure, the
fabrication, and the optical characterization. We also address the
cavity mode spacing by a simple model and finally conclusions are
given.

\section{The resonator structure}

\begin{figure}[b!]
\begin{center}
\epsfig{file=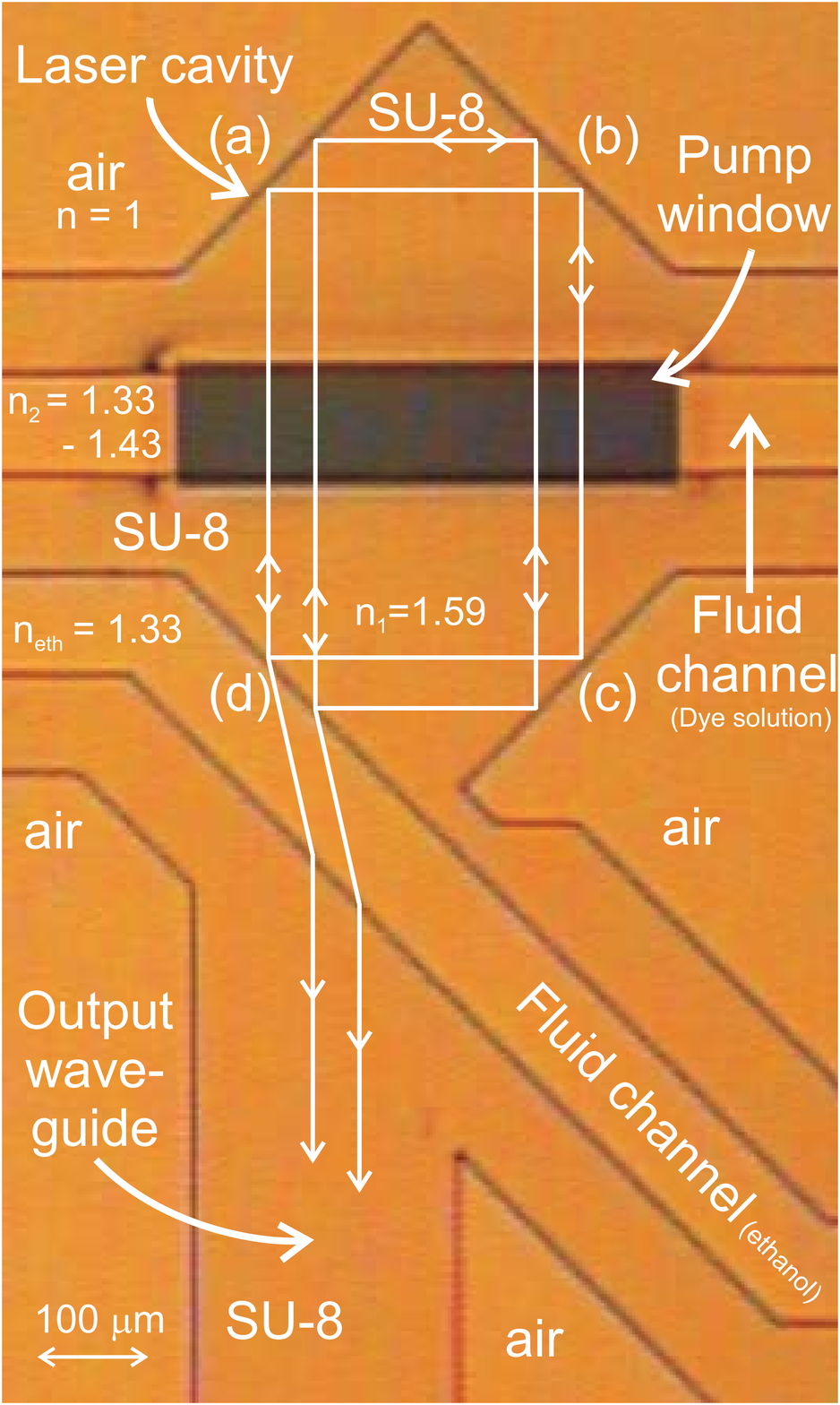, width=0.48\textwidth,clip}
\end{center}
\caption{Optical microscope image of the laser resonator structure
consisting of two isosceles triangles of the photo definable
polymer SU-8 ($n_1 = 1,59$) and a micro fluidic channel
in-between. Two classical trajectories of equal optical path
length are drawn. The optical pumping is performed through a
window (dark rectangular area in the photo) in a metal mask.}
\label{fig:sample}
\end{figure}

The laser resonator, see Fig.~\ref{fig:sample}, resembles a
classical Fabry--Perot resonator. A $\rm{100~\mu m}$ wide
$\rm{10~\mu m}$ deep micro-fluidic channel is bounded by two
isosceles triangle mirrors formed in SU-8, giving rise to the
laterally emitting ring laser cavity. This optical resonator
relies on total internal reflection at the vertical sidewalls (a),
(b), and (c) of the triangles, see Fig.~\ref{fig:sample}. The
laser dye gain medium is located in the micro-fluidic channel
passing through the cavity.

Out-coupling of the laser light is achieved by a second
micro-fluidic channel filled with ethanol, thereby removing the
total internal reflection at the side (d) in the cavity. The
out-coupled light is collected through an integrated planar
'output waveguide' (also in Fig.~\ref{fig:sample}). The output
coupling is switchable by alternating the content of the second
fluidic channel between ethanol and air. A metal mask with a
transparent window at the resonator serves to localize the optical
pumping light to the resonator and diminish unwanted fluorescence.

\section{Fabrication}

\begin{figure}[b!]
\begin{center}
\epsfig{file=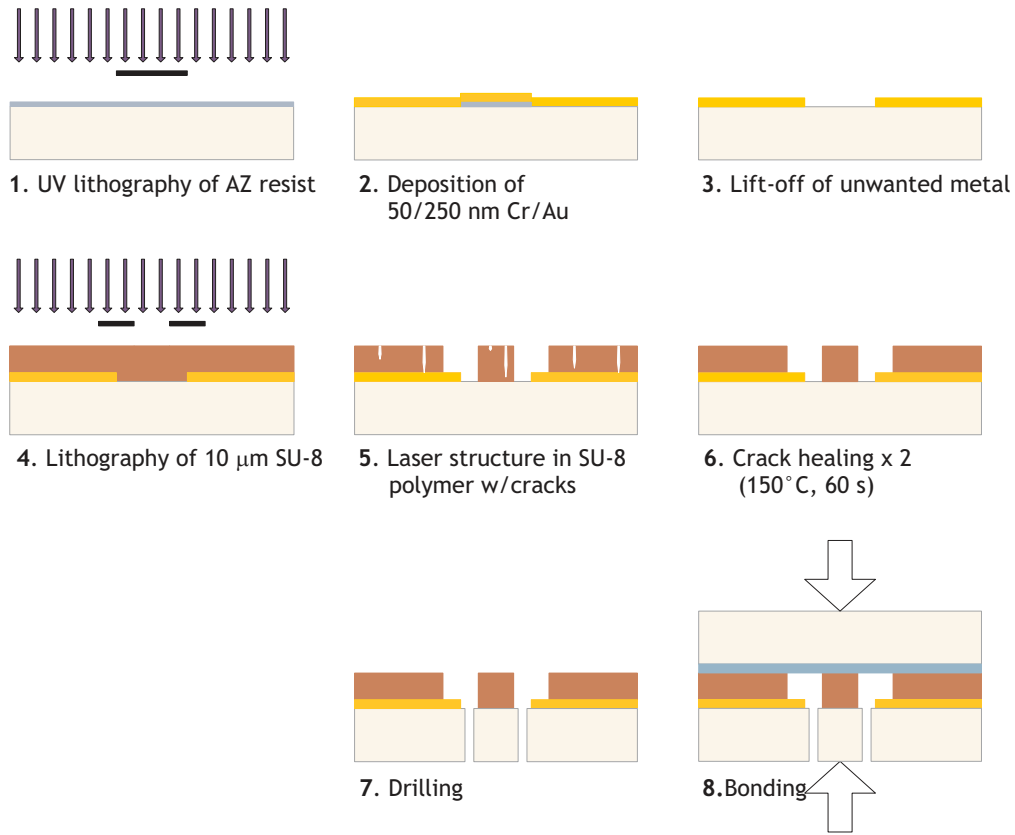, width=0.96\textwidth,clip}
\end{center}
\caption{A schematic of the fabrication sequence. Part 1-3:
Lift-off process. The metal layer is used for screening the
pumping light from certain areas of the chip to avoid unwanted
fluorescence. Part 4-8: SU-8 lithography, drilling, and bonding.}
\label{fig:fabric}
\end{figure}

The fabrication sequence is schematically shown in
Fig.~\ref{fig:fabric}. The process consists of two parts: Step 1-3
is a lift-off process to deposit the metal mask, 50~nm Cr and
250~nm Au, on a glass substrate, and step 4-8 is the forming of
the micro-structures in SU-8, drilling, and bonding a glass lid on
top.

In step 4 the micro fluidic laser structure, the integrated
waveguide, and the micro-fluidic channels (shown in
Fig.~\ref{fig:sample}) are defined by UV lithography on the
substrate in a 10 $\mu$m thick layer of SU-8 photo
resist~\cite{microchem}. Subsequent to development (step 5), the
substrate is placed on a 150$^\circ$C hotplate for $\rm{2 \times
60~s}$ which has the effect of healing the cracks in the SU-8
(step 6). After drilling the holes for the fluid inlets and
outlets (step 7) the micro-fluidic channels are sealed in step 8
by bonding a glass lid on top by means of a 5 $\mu$m thick PMMA
film~\cite{bilenberg2004a}. The bonding is carried out at a
temperature of $120^\circ$C with a bonding force of 2~kN on a 4
inch wafer pair with a duration of 10 minutes.

\section{Optical characterization}

The laser structure is characterized using the laser dye Rhodamine
6G dissolved in i) ethanol and ii) ethylene glycol as the active
gain medium, which is infused at 50~$\rm{\mu L/h}$ through the
micro-fluidic channel and laser resonator. The dye laser is
optically pumped normal to the chip plane through the window (see
Fig.~\ref{fig:sample}) at 532 nm by a pulsed, frequency doubled
Nd:YAG laser. The repetition rate is 10 Hz and the pulses have a
duration of 5 ns.

\begin{figure}[b!]
\begin{center}
\epsfig{file=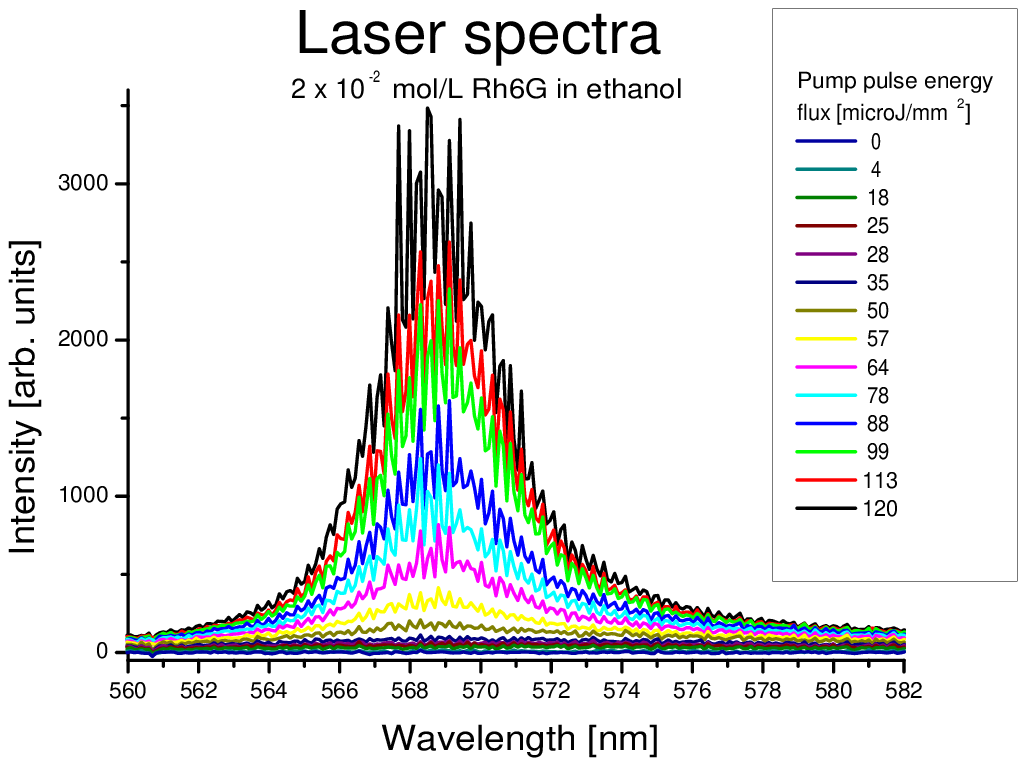, width=0.48\textwidth,clip}
\hspace{5cm} \epsfig{file=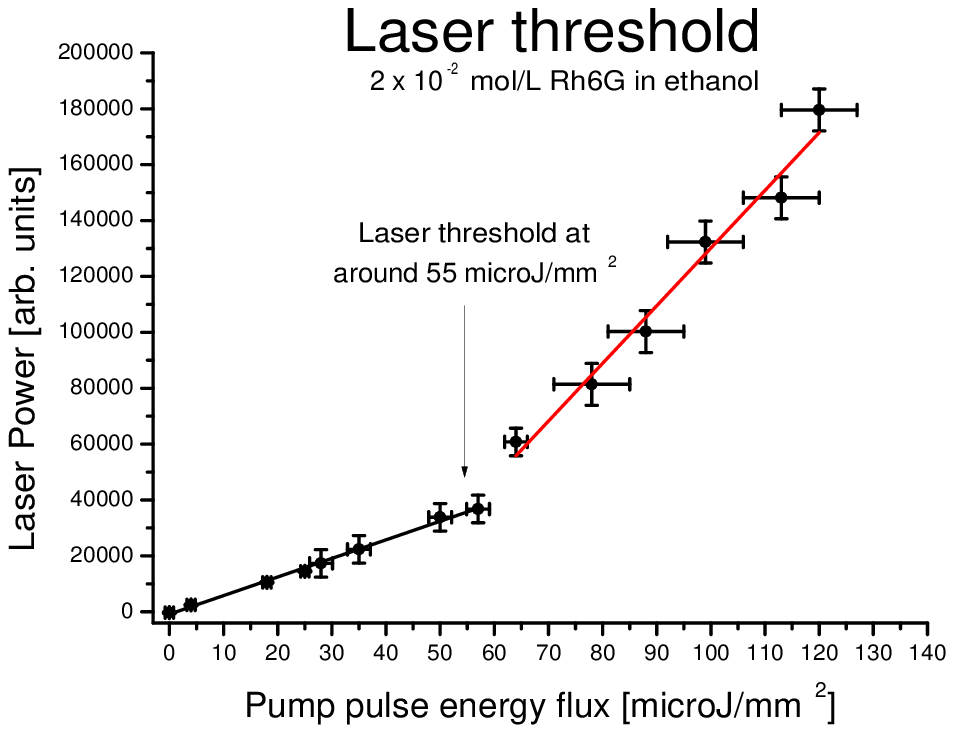,
width=0.48\textwidth,clip}
\end{center}
\caption{Upper panel: Measured laser spectra using Rh6G dissolved
in ethanol ($2 \times 10^{-2}\,{\rm mol/L}$) as the active gain
medium at different pump pulse energy flux values. The dye laser
is pumped at 532 nm using an external frequency doubled Nd:YAG
laser. Lower panel: The laser power vs. the pump pulse energy flux
for the spectra above. To obtain the laser power the optical
output spectra have been numerically integrated from 550-600 nm.
The linear dependence and change in slope reveals laser action.
The lasing threshold is at about 55 $\mu$J/mm$^2$.}
\label{fig:etha_spec}
\end{figure}

\begin{figure}[b!]
\begin{center}
\epsfig{file=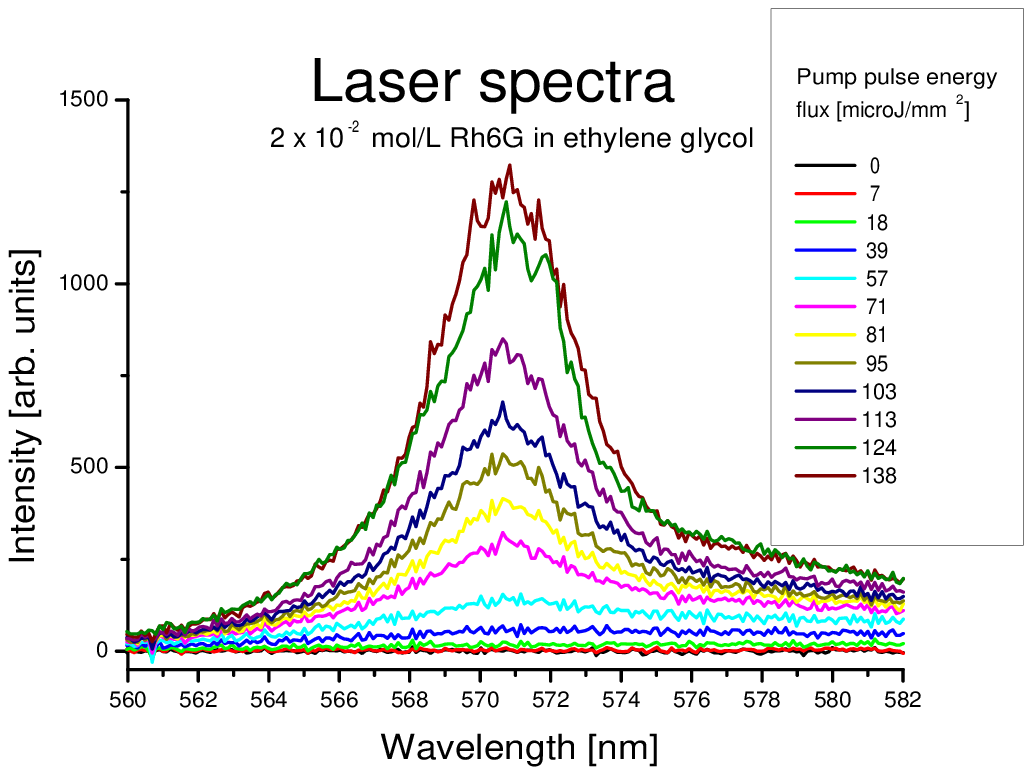, width=0.48\textwidth,clip}
\hspace{5cm} \epsfig{file=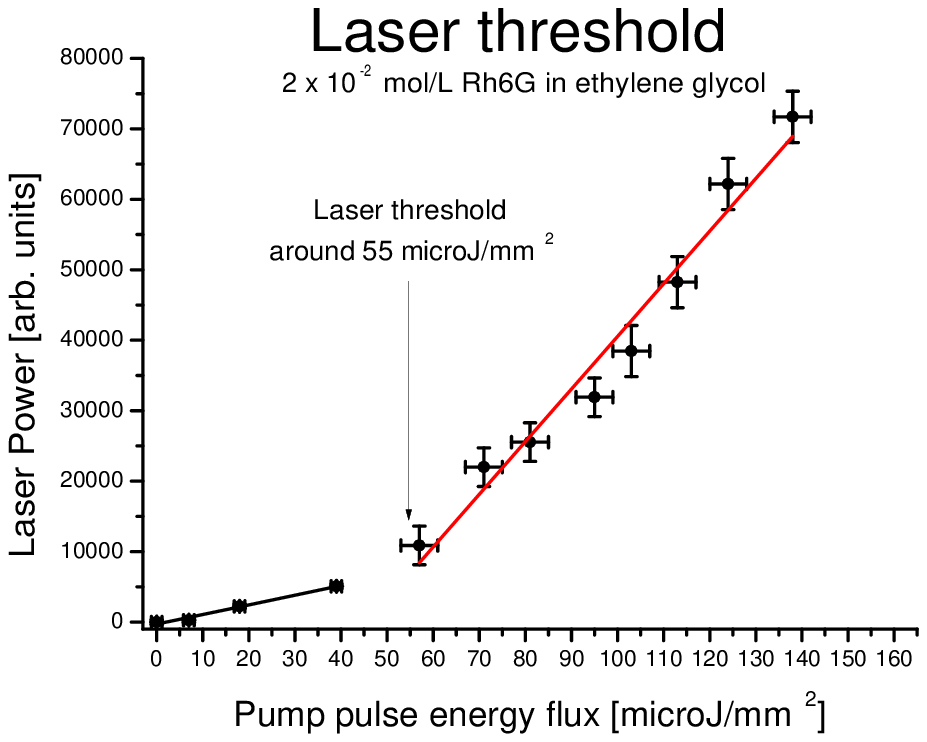,
width=0.48\textwidth,clip}
\end{center}
\caption{Upper panel: Measured laser spectra using Rh6G dissolved
in ethylene glycol ($2 \times 10^{-2}\,{\rm mol/L}$) as the active
gain medium at different pump pulse energy flux values. The dye
laser is pumped at 532 nm using an external frequency doubled
Nd:YAG laser. Lower panel: The laser power vs. the pump pulse
energy flux for the spectra above. To obtain the laser power the
optical output spectra have been numerically integrated from
566-576 nm. The linear dependence and change in slope reveals
laser action. The lasing threshold is at about 55 $\mu$J/mm$^2$.}
\label{fig:ethy_spec}
\end{figure}

The upper panels in Figs.~\ref{fig:etha_spec} and
\ref{fig:ethy_spec} show typical output spectra with the dye
dissolved in ethanol and ethylene glycol, respectively. The
concentration is in both cases $2\times 10^{-2}$~mol/L. By
increasing the pump pulse energy we observe an increase in the
output intensity. The lower panels show the output power
(numerically integrated output intensity) versus the pump pulse
energy flux and the observed change of slope in the curves around
55~${\rm \mu J/mm^2}$ is a clear signature of the onset of lasing
for both solvents.

\begin{figure}[t!]
\begin{center}
\epsfig{file=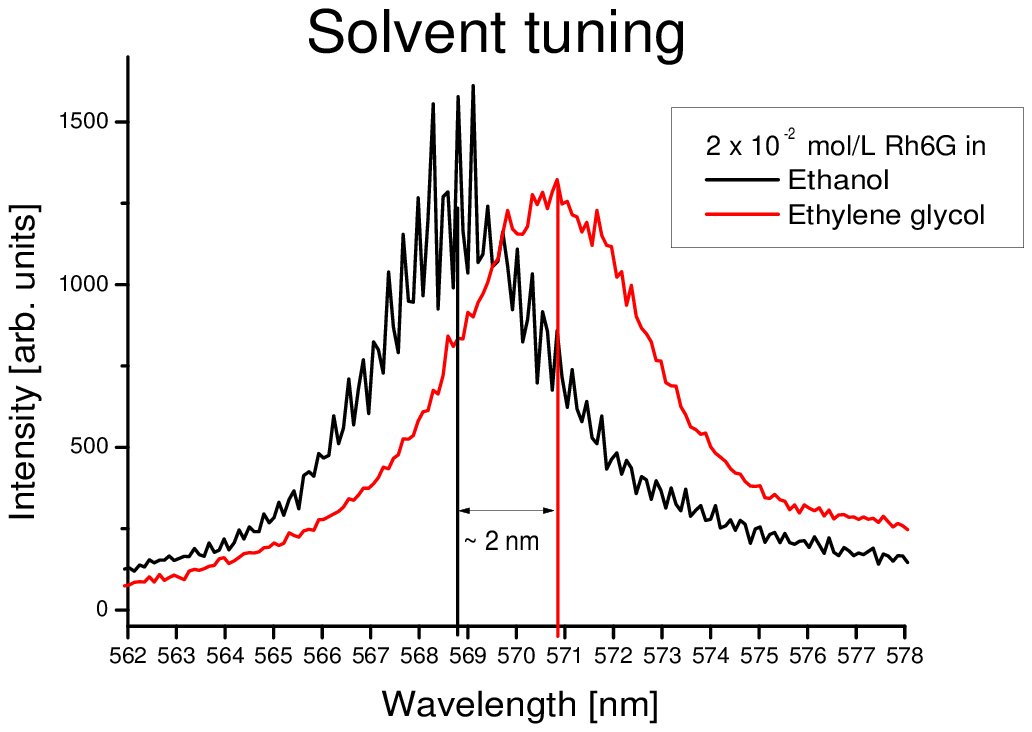, width=0.48\textwidth,clip}
\end{center}
\caption{Two measured laser spectra using Rh6G dissolved in
ethanol and ethylene glycol, respectively, as the active gain
medium (same concentration). The dye laser is pumped at 532 nm
using an external frequency doubled Nd:YAG laser. The change of
solvent provides tuning capabilities of the laser due to a shift
of the gain-spectrum of the dissolved laser dye. A red shift of
approx. 2 nm is observed when changing from ethanol (refractive
index $n=1.33$) to ethylene glycol ($n=1.43$).}
 \label{fig:solvent}
\end{figure}

As shown in Fig.~\ref{fig:solvent}, the change of solvent
introduces an over-all shift of the output spectra of
approximately 2~nm. The red shift of the spectra measured using
the ethylene glycol solution may be explained by a lower
cavity-loss or a higher dye quantum efficiency using this solvent
\cite{peterson1971}. Since the dye/solvent properties can be
changed dynamically this enables real-time tuning of the central
lasing wavelength \cite{bilenberg2003a}.

The output coupling from the structure is switched on by filling
the second fluid channel (see Fig.~\ref{fig:sample}) with ethanol,
and it is switched off again by blowing air into the second
channel or providing an under-pressure with a syringe, thereby
sucking air through the second channel.

\section{The cavity modes}

\begin{figure}[b!]
\begin{center}
\epsfig{file=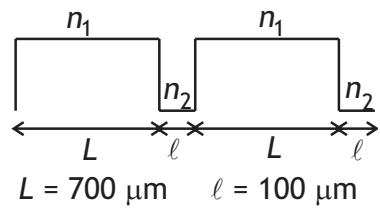, width=0.35\textwidth,clip}
\end{center}
\caption{For a calculation of the cavity mode spectrum the
structure can effectively be 'unfolded'. The sketch illustrates
the variation of the refractive index for one round-trip in the
cavity depicted in Fig.~\ref{fig:sample}. $n_1=1.59$,
$n_2=1.33-1.43$. } \label{fig:unfold}
\end{figure}

The cavity mode spectrum has been modelled by a scattering matrix
approach and Fig.~\ref{fig:unfold} illustrates the 'unfolded'
structure. Formally, the scattering matrix of the 'unfolded'
structure is calculated and periodic boundary conditions are
applied. In other words the ring resonator is formed by 'joining'
the two ends of the unfolded structure. This allows for
calculation of the spectral position of the cavity modes. For the
experimentally relevant parameters a mode spacing of approximately
0.13 nm is found.

Comparing to the experimental output spectra in
Figs.~\ref{fig:etha_spec} and~\ref{fig:ethy_spec} we might
speculate that the sub-nanometer modulation of the overall peak
resembles the longitudinal mode spacing. However, since the
spectrometer resolution is around 0.15 nm the mode-spacing is not
resolved experimentally. Compared to the estimated mode-spacing
the over-all width of the output spectra suggests that the lasing
is longitudinally multi-mode.

\section{Conclusion}

We have demonstrated a laterally emitting coupled cavity
micro-fluidic dye ring laser. The laser is fabricated with polymer
technology using a total-internal reflection based cavity formed
by two triangle mirrors. The light-source is straight forward to
integrate with polymer planar waveguides for use in lab-on-a-chip
devices. The dependence of the lasing wavelength on the dye
solution may enable tunability. Using a simple modelling approach
we estimated the mode-spacing of the order 0.13~nm which however
can not be fully verified experimentally due to limited
spectrometer resolution. We have shown the possibility of
switching the output coupling which can be highly useful also in
other applications.

\section*{Acknowledgements}

The first author gratefully acknowledges financial support by Otto
M\o nsteds Fond and Geheimestatsminister Greve Joachim Godske
Moltkes Legat from Bregentved Gods.

%\bibliographystyle{elsart-num}
%\bibliography{cavity}

\end{document}